\documentclass{article}
\usepackage{listings}
\usepackage{xcolor}
\usepackage{amsmath}
\usepackage{amssymb}
\usepackage{stmaryrd}
\usepackage{url}

\usepackage{graphicx}

\newcommand{\bR}{\mathbb{R}}
\newcommand{\lift}{\text{lift}}

\definecolor{codegreen}{rgb}{0,0.6,0}
\definecolor{codegray}{rgb}{0.5,0.5,0.5}
\definecolor{codepurple}{rgb}{0.58,0,0.82}
\definecolor{backcolour}{rgb}{0.95,0.95,0.92}

\lstdefinestyle{mystyle}{
    language=Python,
    backgroundcolor=\color{backcolour},   
    commentstyle=\color{codegreen},
    keywordstyle=\color{magenta},
    numberstyle=\tiny\color{codegray},
    stringstyle=\color{codepurple},
    basicstyle=\ttfamily\footnotesize,
    breakatwhitespace=false,         
    breaklines=true,                 
    captionpos=b,                    
    keepspaces=true,                 
    numbers=left,                    
    numbersep=5pt,                  
    showspaces=false,                
    showstringspaces=false,
    showtabs=false,                  
    tabsize=2
}

\begin{document}

\title{Lifting E-Graphs: A Function Isn't a Constant}

\author{Philip Zucker}

\maketitle
\begin{abstract}
Variables are quite subtle and easy to get wrong. An approach is described to support rigid $\alpha$ canonical variables in an e-graph. The lifting e-graph has a baked-in notion of functional lifting combinator. It is implemented by fattening the usual integer identifiers with thinning bitvectors, lift-pulling smart constructors, and a special thinning-aware union find variation. The approach is inspired by slotted e-graphs \cite{slotted} and Co-de Bruijn syntax \cite{McBride_2018}.
\end{abstract}

\section{Introduction}
An E-graph \cite{egg} is a data structure that compactly store ground terms and equalities between them.

E-graphs are proposed to be useful for program optimization \cite{PEG}. Program expressions such as  $x * 2 / 2$ usually have unknown variables in them.

As a first approach, these unknowns are usually referred to by named variables. However, there are a number of issues with explicit names like $x$ in the e-graph setting

\begin{enumerate}
\item Generative processes like equality saturation can bloat the e-graph with many redundant, essentially equivalent expressions if fresh names are created via a gensym, for example in the rewrite rule $P \rightarrow \forall x, P$ or $e \rightarrow \frac{1}{N} \sum_{i=0}^{N-1} e$
\item The equality in an e-graph has an unclear relationship with scope hygiene. For example, the equality $x * 0 = 0$ can rewrite $0$ to cause $x$ to appear in $\text{lam}(y, 0) = \text{lam}(y,x * 0)$ which does not have $x$ bound.
\item Named variables have missed memory sharing opportunities. $\texttt{f(g(h(x)))}$ shares no storage with $\texttt{f(g(h(y)))}$. The amount of missed storage opportunity gets worse the deeper and bigger the term. Note that these two terms \emph{aren't equal}, but there is a missed \emph{relationship} that can be used for reasoning and compaction.
\item Depending on your intended semantics, expressions like $\sin(x)$ may ambiguously refer to multiple semantic entities. In other words, there may be \emph{too much} sharing. 
\end{enumerate}

\subsection{Which $\sin$ is $\sin(x)$?}

$\sin(x)$ from an ordinary perspective is fine. We've lived with the notation for hundreds of years. It works. We know what it means when we see it.

From another perspective it is horribly vague and possibly collapsing distinct entities. We are somehow referring to the function $\sin$, but are we specifically referring to $x \mapsto \sin(x) : R^1 \to R$ or to $x,y \mapsto \sin(x) : R^2 \to R$?

\begin{figure}[htbp]
\centering
\includegraphics[width=0.5\linewidth]{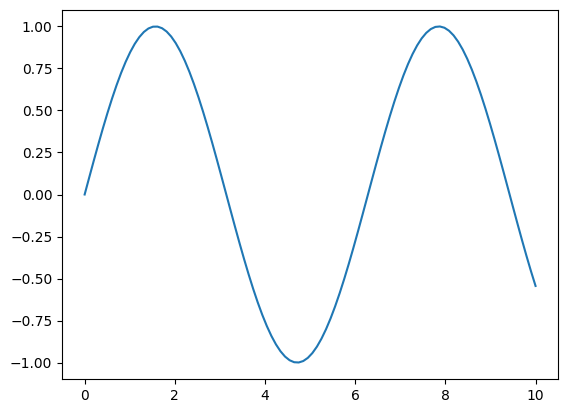}
\caption{$x \mapsto \sin(x) : R \to R$}
\label{fig:sin-r-to-r}
\end{figure}

\begin{figure}[htbp]
\centering
\includegraphics[width=0.5\linewidth]{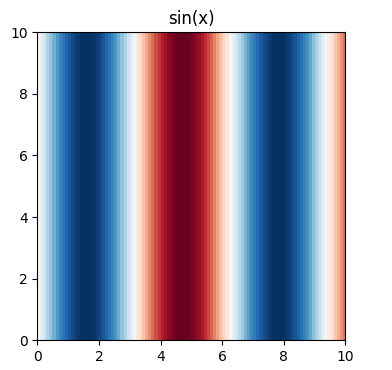}
\caption{$x,y \mapsto \sin(x) : R^2 \to R$}
\label{fig:sin-r2-to-r}
\end{figure}

These are different things. In fact, they can't even be compared for equality due to type mismatch. The two plots are completely different images. Yet, the notation that suppresses the context $x,y \mapsto \_$ or leaves it implicit conflates these two distinct things. It is at least worrying that perhaps in some subtle way this conflation is asserting $\bR^2 = \bR^1$ and from thence unsoundness ensues.

While it is not the case for every conception of the word "context" or the word "term", from this observation comes a design philosophy:

\textbf{The context $x,y \mapsto \_$ is not \emph{where} a term is, it is part of \emph{what} a term is}

\section{Naive Well-Dimensioned Nameless Representation}

It is helpful to find a way to talk about this less familiar $x,y \mapsto \_$  notation in the more standard framework of first-order terms.

If there is an ambiguity about dimension, a simple approach to break this ambiguity is to add a $d$ dimension subscript on all function symbols.

We can make a different copy of every function symbol for every dimension/context we might be working in and we can refer to variables by (dimension,index) pairs $\text{var}_{di}$ rather than by names. For example, $x \mapsto x$ becomes $\text{var}_{10}$ (the zeroth variable in context of size 1) and $x,y \mapsto y$ becomes $\text{var}_{21}$ (the first variable in context of size 2).

Likewise, we could also disambiguate all the $\sin$ into different versions $\sin_d$ depending on the type of its argument. If $\text{var}_{21}$ has type $\bR^2 \rightarrow \bR$ then if $\sin$ is going to accept it, it needs to take in arguments of that type. We have $\sin_0 : (\bR^0 \rightarrow \bR) \rightarrow (\bR^0 \rightarrow \bR)$, $\sin_1 : (\bR^1 \rightarrow \bR) \rightarrow (\bR^1 \rightarrow \bR)$,  $\sin_2 : (\bR^2 \rightarrow \bR) \rightarrow (\bR^2 \rightarrow \bR)$ and so on.

There is a pointwise compositional semantics of these combinators. 

\[
\begin{aligned}
\llbracket 42_d \rrbracket &= v_0, v_1, \ldots, v_{d-1} \mapsto 42 \\
\llbracket \operatorname{var}_{di} \rrbracket &= v_0, v_1, \ldots, v_{d-1} \mapsto v_i \\
\llbracket \sin_d(t) \rrbracket &= v_0, v_1, \ldots, v_{d-1} \mapsto \sin(\llbracket t \rrbracket(v_0, v_1, \ldots, v_{d-1})) \\
\llbracket t +_d s \rrbracket &= v_0, v_1, \ldots, v_{d-1} \mapsto \llbracket t \rrbracket(v_0, v_1, \ldots, v_{d-1}) + \llbracket s \rrbracket(v_0, v_1, \ldots, v_{d-1}) \\
&\text{and so on...}
\end{aligned}
\]

Semantically, $\operatorname{var}_{di}$ is a \emph{projection} of type $\mathbb{R}^d \rightarrow \mathbb{R}$. $\sin_d$ is a \emph{higher order function} of type $(\mathbb{R}^d \rightarrow \mathbb{R}) \rightarrow (\mathbb{R}^d \rightarrow \mathbb{R})$. It can also be viewed as the partial application of the composition function $\sin_d \sim \sin \circ \_$.

Because we are being nameless by referring to variables by integers, $x \mapsto f(g(h(x)))$ becomes syntactically the same as $y \mapsto f(g(h(y)))$ since both become $f_1(g_1(h_1(\text{var}_{10})))$. So sharing in the e-graph using this representation has become better in that sense.

On the other hand, $\sin_1(\text{var}_{10}) : \mathbb{R}^1 \rightarrow \mathbb{R}$ and $\sin_2(\text{var}_{20}) : \mathbb{R}^2 \rightarrow \mathbb{R}$ now share no storage. This is fine from the perspective that they are not \emph{equal} (such a question does not even type check), but is disappointing from the perspective that they are clearly \emph{related}.

\section{Lifting Functions for Increased Sharing}
Lifting is a functional combinator that drops some arguments and passes others along while maintaining their ordering. 

One way that $\sin_1(\text{var}_{10}) : \mathbb{R}^1 \rightarrow \mathbb{R}$ and $\sin_2(\text{var}_{20}) : \mathbb{R}^2 \rightarrow \mathbb{R}$ are related is by this operation of $\text{lift}_t : (\mathbb{R}^{\text{popcnt}(t)} \rightarrow \mathbb{R})\rightarrow (\mathbb{R}^{\text{len(t)}} \rightarrow \mathbb{R})$. For example $\text{lift}_{110} : (\mathbb{R}^{2} \rightarrow \mathbb{R})\rightarrow (\mathbb{R}^3 \rightarrow \mathbb{R})$  keeps arg 0, keeps arg 1, and drops arg 2.

It can be written as a simple Python combinator:

\begin{lstlisting}
def lift(thin : Thin):
    return lambda f: lambda *args: f(*act(thin, args))
\end{lstlisting}

Using this concept, we can make the differently dimensioned $\sin$ expressions share subterms and hence achieve shared memory storage.

$$\sin_1(\text{var}_{10}) := \sin(\text{var})$$

$$\sin_2(\text{var}_{20}) := \text{lift}_{10}(\sin(\text{var}))$$

\begin{figure}[htbp]
\centering
\includegraphics[width=0.5\linewidth]{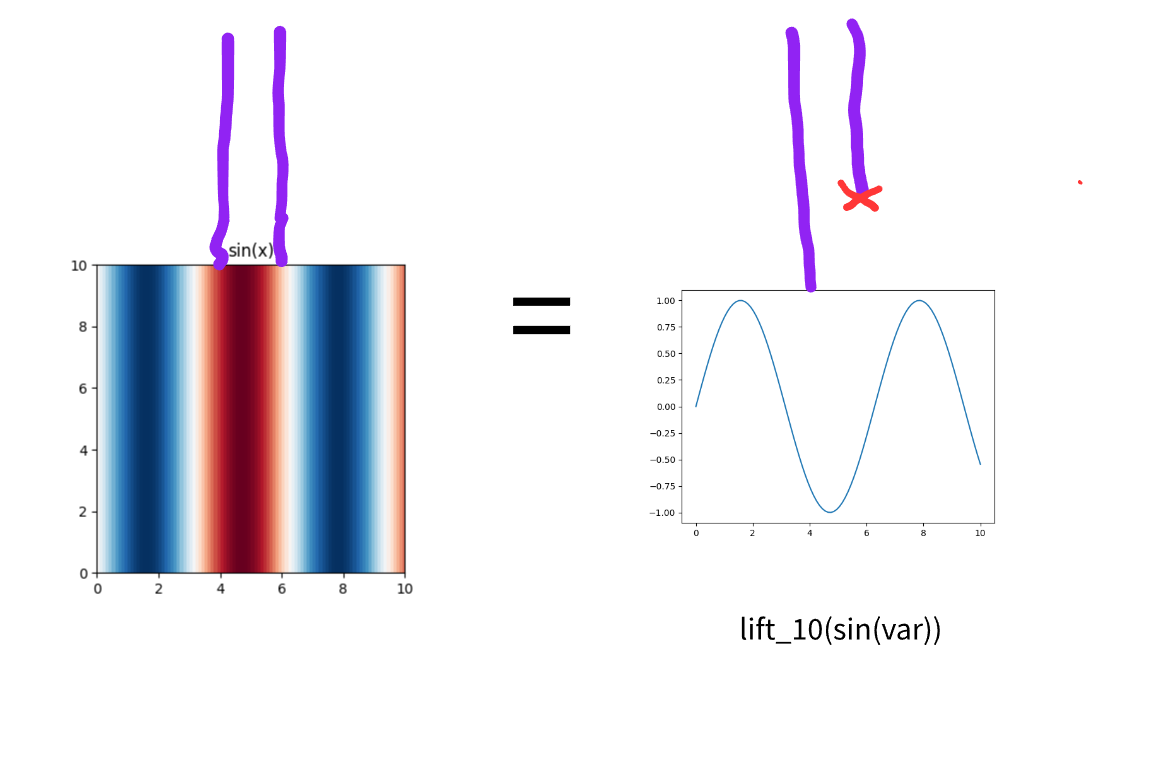}
\caption{A lifted sine expression}
\label{fig:thin-sin-sin}
\end{figure}


The lifting is parametrized by a thinning \cite{McBride_2018}\cite{zucker2026thinningssub} bitvector $t$ with a $1$ for arguments to keep and $0$ for arguments to drop. One \emph{lifts} a function by \emph{thinning} its arguments list.

Thinnings can be thought of in a number of ways. They are strictly monotonic maps between totally ordered finite sets. They can also be seen as a recipe to extract subsequences.

They are similar mathematical objects to permutations, albeit more minimalist in data requirements and implementations. They can be seen as a compaction of multiple de Bruijn shifts in a manner similar to how a permutation dictionary is a compaction of many swaps \cite{zucker2026thinningssub}. When a de Bruijn shift occurs while pushing an expression inside of new binders, a gap is created between the indices referring to variables inside the expression and the indices referring to variables at the expression's original context. When one creates these gaps multiple times, one has created something akin to a thinning.

Thinnings have a notion of composition and identity. This makes them a category. They are relatives of the simplex category.


\begin{figure}[htbp]
\centering
\includegraphics[width=0.5\linewidth]{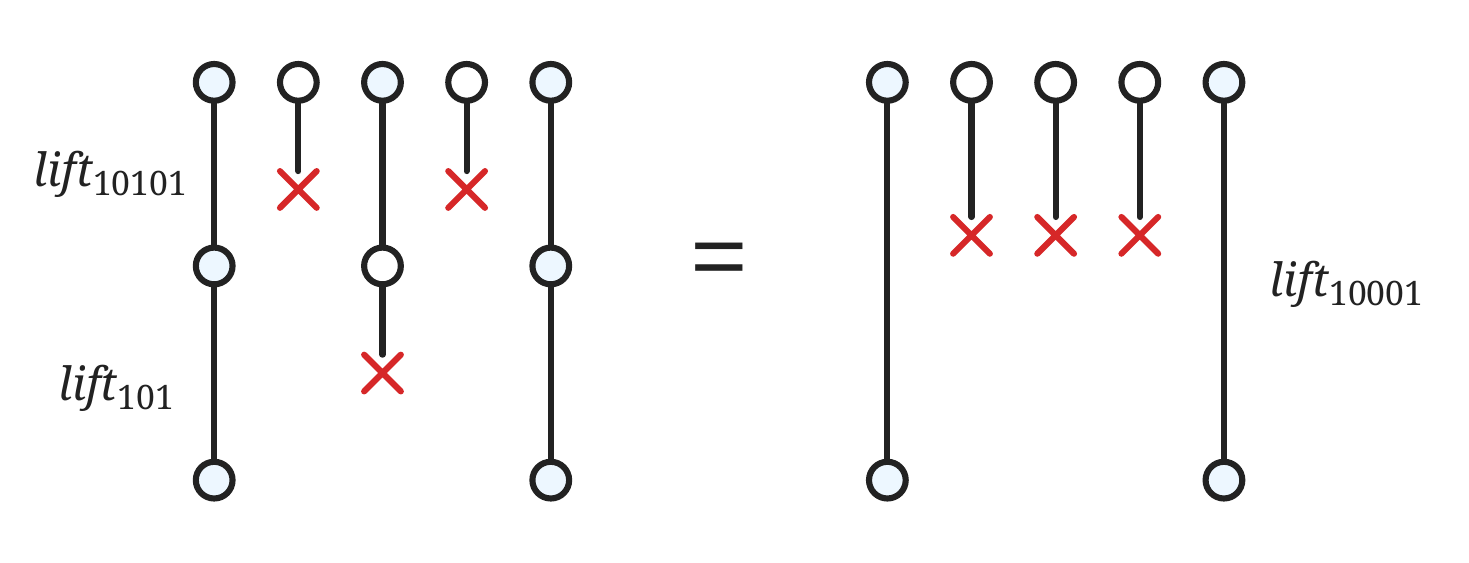}
\caption{Composition of thinnings}
\label{fig:compose-thinnings}
\end{figure}

\begin{lstlisting}
type Thin = list[bool]
def dom(f : Thin) -> int:
    return len(f)
def cod(f : Thin) -> int:
    return sum(f)
def id(n : int) -> Thin:
    return [True]*n
def comp(f : Thin, g : Thin) -> Thin:
    assert cod(f) == dom(g)
    i = 0 
    result = []
    for a in f:
        if a:
            result.append(g[i])
            i += 1
        else:
            result.append(False)
    assert i == len(g)
    return result
\end{lstlisting}


Lifting has some useful equational properties. The "parametric polymorphism" of typical pointwise derived combinators like $\sin$ manifests as a rewrite rule $\sin(\lift_i(X)) = \lift_i(\sin(X))$ . This is stating that $\lift$ is a homomorphism with respect to typical function symbols. The operations of adding redundant arguments and applying a function like $\sin$ pointwise commute. Syntactically, it allows for pushing and pulling common $\lift$ expressions through function symbols.


Thinning composition manifests as a constant propagation rule for liftings $\lift_i(\lift_j(X)) = \lift_k(X)$ where $k = i \cdot j$.

\begin{enumerate}
\item $\forall X, \lift_i(\lift_j(X)) = \lift_{i \cdot j}(X)$ lift compaction rules
\item $\forall X Y, f(\lift_i(X), \lift_i(Y)) = \lift_i(f(X,Y))$ lift pulling rules
\end{enumerate}

\section{Baking Lift In}

These properties of lifting are so simple, ubiquitous and structural that it may make sense to bake them into the very fabric of what a term or an egraph \emph{is}. This is sweetened by the fact that liftings/thinnings can be represented as compact bitvectors \cite{gonccalves2025thinning}.

An approach to create an E-Graph Modulo Theories (EMT) \cite{zucker2025omeletsneedonionsegraphs} is to use fat identifiers that contain thinnings. By making the lifting a part of the identifier, it becomes ephemeral instead of interned and immediately available for inspection when needed.

\begin{lstlisting}
type FatId = (list[bool], int)
\end{lstlisting}

\subsection{Lift Pulling Smart Constructors}

The ordinary e-node datatype now has children that are \lstinline|FatId|.

\begin{lstlisting}
class ENode:
  f : str
  args : list[FatId]
\end{lstlisting}

The homomorphism rules can be oriented to pull the liftings as high as possible  $f(\lift_i(X), \lift_i(Y)) \to \lift_i(f(X,Y))$, leaving behind only the minimal thinnings needed to reconcile the contexts of the arguments of a function symbol.


Whenever you build a new e-node, the smart constructor should examine the common lifting of the fat eids of its arguments, peel off this lifting, intern the e-node, and then put the common lifting back on before returning the fat eid to the user. This is a operational way of achieving the lift pull-up rule inside the e-graph.

The smart constructor operation will ensure that whenever you build a node with arguments e-ids that are lifted more than necessary, you get back a fat e-id handle to the same interned data regardless of the extra lifting, enabling reduced memory usage and faster lifting relationship comparison. Overly lifted e-ids may appear for the convenience of a user to work in a large context but may also appear during the rebuilding phase of an e-graph.


Even in the absence of a union find, this lifting pulling smart constructor with fat ids makes for an $\alpha$ aware hash cons. \cite{zucker2026alphaequival} \cite{hashalpha}

Pulling lifts up corresponds in an interesting way to the co-de Bruijn style of normalizing and representing lambda terms as described in McBride's Everybody's Got to Be Somewhere \cite{McBride_2018} . 

Note also that by being as thin as possible, the dimensionality can play the role of a kind of a nameless free variable analysis. Semantically speaking, the ability to be written in lifted form is stating that the entity in question is known to be constant in the directions that have been dropped. As equality saturation progresses, one may learn that the function in question depends on fewer directions than previously thought. By being part of the fabric of the term's semantics, it is less of a problem to make sure that the more syntactically flavored notion of free variable analysis is up to date.

\begin{table}[htbp]
\centering
\small
\renewcommand{\arraystretch}{1.4}
\begin{tabular}{p{0.25\linewidth}p{0.34\linewidth}p{0.34\linewidth}}
\hline
Named Form & Unnormalized & Max Pulled \\
\hline
$x,y,z,w \mapsto 42$
&
$\lift_{0000}(42)$
&
$\lift_{0000}(42)$
\\
$x,y,z,w \mapsto 42 + 42$
&
$\lift_{0000}(42) + \lift_{0000}(42)$
&
$\lift_{0000}(42 + 42)$
\\
$x,y,z,w \mapsto z + 42$
&
$\lift_{0010}(\operatorname{var}) + \lift_{0000}(42)$
&
$\lift_{0010}(\operatorname{var} + \lift_0(42))$
\\
$x,y,z,w \mapsto x + z$
&
$\lift_{1000}(\operatorname{var}) + \lift_{0010}(\operatorname{var})$
&
$\begin{aligned}
\lift_{1010}(&\lift_{10}(\operatorname{var}) \\
&+ \lift_{01}(\operatorname{var}))
\end{aligned}$
\\
\hline
\end{tabular}
\caption{Named forms and pulled lifting expressions}
\label{tab:named-forms}
\end{table}

\subsection{Thinning Union Find}

But we want an e-graph. We need to add a union find to that hash cons.

\begin{lstlisting}
type UnionFind = list[FatId]
\end{lstlisting}

How do we implement a union find that accepts lifted fat eids? It is easiest to explain the needed extra capabilities (annotation force parent picking and make-set) but examining some simpler asymmetric annotated union finds first.

\subsubsection{Offset (Group) Union Finds}

One may want to add extra information in the union find, either on the edges, on the nodes, or on the roots / classes.

One thing you can do at little cost is express offset relationships like $e_4 + 3 = e_8 - 4$.  We can do this by adding an offset annotation to the parents table and collecting up the offsets as we traverse up the tree. \cite{groupuf} \cite{offset_congruence}

\begin{lstlisting}
from dataclasses import dataclass, field
type Id = tuple[int, int] # offset and id
@dataclass
class OffsetUF:
    parents : list[Id] = field(default_factory=list)
    def makeset(self) -> Id:
        i = len(self.parents)
        self.parents.append((0,i))
        return (0,i)
    def find(self, x : Id) -> Id:
        off, xid = x
        while True:
            (offy, yid) = self.parents[xid]
            if yid == xid:
                assert offy == 0
                return (off, xid)
            off += offy
            xid = yid
    def union(self, x : Id, y : Id) -> Id | None:
        (offx, xid) = self.find(x)
        (offy, yid) = self.find(y)
        if xid != yid:
            z = (offy - offx, yid)
            self.parents[xid] = z
            return z
        else:
            return None
\end{lstlisting}

\subsubsection{Monus Union Find}

A slight but interesting twist is to require the offset annotations only be positive. To achieve this, the annotations now need control of the direction union adds to the parents table.

\begin{lstlisting}
    def union(self, x : Id, y : Id) -> Id | None:
        (offx, xid) = self.find(x)
        (offy, yid) = self.find(y)
        if xid != yid:
            if offy >= offx:
                z = (offy - offx, yid)
                self.parents[xid] = z
            elif offx > offy:
                z = (offx - offy, xid)
                self.parents[yid] = z
            return z
        else:
            return None
\end{lstlisting}

\subsubsection{Factor Union Find}

Another slight twist on the above is to consider multiplication by integer constants 6*x = 4*y. Because we don’t insist on turning multiplication of $\mathbb{Z}$ into a group, it isn’t clear we can derive a solution in the form $x = ?a * y$ or $y = ?b * x$ without constraining more than is implied by the assertion. Nevertheless there is a “best” solution if we create a new identifier $x = 2*z$, $y = 3*z$. These numbers come out of considerations of the prime factorization / least common multiple. The right hand side says there must be at least 2 factors of 2, but 6 only has 1 factor, so we must pull another 2 out of x. Likewise we must pull a factor of 3 out of $y$.

In short, union not only controls the directionality of the parent, it also must sometimes generate a fresh id.

\begin{lstlisting}
    def union(self, x : Id, y : Id) -> Id | None:
        (mx, xid) = self.find(x)
        (my, yid) = self.find(y)
        if xid != yid:
            mz = math.lcm(mx,my)
            if mz == mx:
                z = (mx // my, x)
                self.parents[yid] = z
                return z
            elif mz == my:
                z = (my // mx, y)
                self.parents[xid] = z
                return z
            else:
                (_, zid) = self.makeset()
                self.parents[xid] = (mz // mx, zid) 
                self.parents[yid] = (mz // my, zid)
                return (mz, zid)
        else:
            assert mx == my
            return None

\end{lstlisting}

\subsubsection{Thinning Union Find}
Because liftings are semantically injective functions, when you union two lifted eids $\lift_i(a) = \lift_i(b)$, you can peel off the common parts of their liftings and learn $a = b$. This is similar to the move you can make in syntactic unification or from an equality between algebraic datatypes, which are also injective functions. $\texttt{cons}(a, c) = \texttt{cons}(b, c)$ implies $a = b$.

If what is left is bare eids $e_6 = e_{47}$ because the two liftings were identical, then it is the ordinary union find action at that point.

It does not even type check to union two objects with different numbers of variables. If a lifting mismatch is due to this, it is a user error.

\subsubsection{Ordinary Union}

\begin{table}[htbp]
\centering
\small
\renewcommand{\arraystretch}{1.4}
\begin{tabular}{p{0.25\linewidth}p{0.68\linewidth}}
\hline
Step & Form \\
\hline
Named Form
&
$(x,y \mapsto y * 1) = (x,y \mapsto y)$
\\
Pulled Form
&
$\lift_{01}(\operatorname{var} * \lift_0(1)) = \lift_{01}(\operatorname{var})$
\\
Interned Form
&
$\lift_{01}(e_{17}) = \lift_{01}(e_{42})$
\\
Peel
&
$e_{17} = e_{42}$
\\
Orient
&
$e_{17} \rightarrow e_{42}$ (or $e_{42} \rightarrow e_{17}$)
\\
\hline
\end{tabular}
\caption{Ordinary union by peeling off a common lift}
\label{tab:ordinary-union}
\end{table}

\subsubsection{$x * 0 = 0$ and Redundancies}

There remain legitimate cases of unequal thinnings being unioned.

A concerning counterexample to any discussion of variables in e-graphs is $x * 0 = 0$. If this is a truly bidirectional equality, it allows $x$ to slip into any location where $0$ is used. 

From the careful scoping/lifting perspective, the actual equation in question is $(x \mapsto x * 0) = (x \mapsto 0)$ which is combinatorized into $\text{var} * \lift_0(0) = \lift_0(0)$. Note that since $0$ is a constant (a 0-arity function $[] \mapsto 0$), it must be lifted into the current 1-context $x \mapsto$.




\begin{table}[htbp]
\centering
\small
\renewcommand{\arraystretch}{1.4}
\begin{tabular}{p{0.25\linewidth}p{0.68\linewidth}}
\hline
Step & Form \\
\hline
Named Form
&
$(x \mapsto x * 0) = (x \mapsto 0)$
\\
Pulled Form
&
$\operatorname{var} * \lift_0(0) = \lift_0(0)$
\\
Interned Form
&
$e_{92} = \lift_0(e_8)$
\\
Orient
&
$e_{92} \rightarrow \lift_0(e_8)$ (forced)
\\
\hline
\end{tabular}
\caption{Forced orientation for a redundant variable}
\label{tab:forced-orientation}
\end{table}

\subsubsection{Make-set Parent}

It is also possible to be in a situation in which neither left nor right side is solvable in terms of the other. This situation is luckily still solvable by generating a common fresh constant that both left and right are solvable to $\texttt{left -> annot1(fresh)}$ and $\texttt{right -> annot2(fresh)}$.




\begin{table}[htbp]
\centering
\small
\renewcommand{\arraystretch}{1.4}
\begin{tabular}{p{0.25\linewidth}p{0.68\linewidth}}
\hline
Step & Form \\
\hline
Named Form
&
$(x,y \mapsto x * 0) = (x,y \mapsto 0 * y)$
\\
Pulled Form
&
$\lift_{10}(\operatorname{var} * \lift_0(0)) = \lift_{01}(\lift_0(0) * \operatorname{var})$
\\
Interned Form
&
$\lift_{10}(e_{92}) = \lift_{01}(e_{13})$
\\
Orient
&
$e_{92} \rightarrow \lift_0(e_{99})$ and $e_{13} \rightarrow \lift_0(e_{99})$ ($e_{99}$ fresh)
\\
\hline
\end{tabular}
\caption{Make-set parent for irreconcilable liftings}
\label{tab:makeset-parent}
\end{table}

\begin{lstlisting}
def wct(f : Thin, g : Thin) -> Thin:
    # weakest common thinning
    assert dom(f) == dom(g)
    return [a and b for a,b in zip(f,g)]

def div(f : Thin, g : Thin) -> Thin:
    assert dom(f) == dom(g)
    assert all(not a for a,b in zip(f,g) if not b) # f is thinner than g
    return [a for a,b in zip(f,g) if b]

type Id = tuple[Thin, int]
@dataclass
class ThinUF:
    parents : list[Id] = field(default_factory=list)
    def makeset(self, scope : int) -> Id:
        i = len(self.parents)
        id = ([True]*scope, i)
        self.parents.append(id)
        return id
    def find(self, x : Id) -> Id:
        thin, xid = x
        while True:
            (thiny, yid) = self.parents[xid]
            if xid == yid:
                assert all(thiny)
                return (thin, xid)
            thin = comp(thiny, thin)
            xid = yid
    def union(self, x : Id, y : Id) -> Id | None:
        thinx, xid = self.find(x)
        thiny, yid = self.find(y)
        if xid != yid or thinx != thiny:
            thinz = wct(thinx,thiny)
            (_, z) = self.makeset(cod(thinz))
            self.parents[xid] = (div(thinz,thinx), z)
            self.parents[yid] = (div(thinz,thiny), z)
            return (thinz, z)
        else:
            return None
\end{lstlisting}

\begin{figure}[htbp]
\centering
\includegraphics[width=0.5\linewidth]{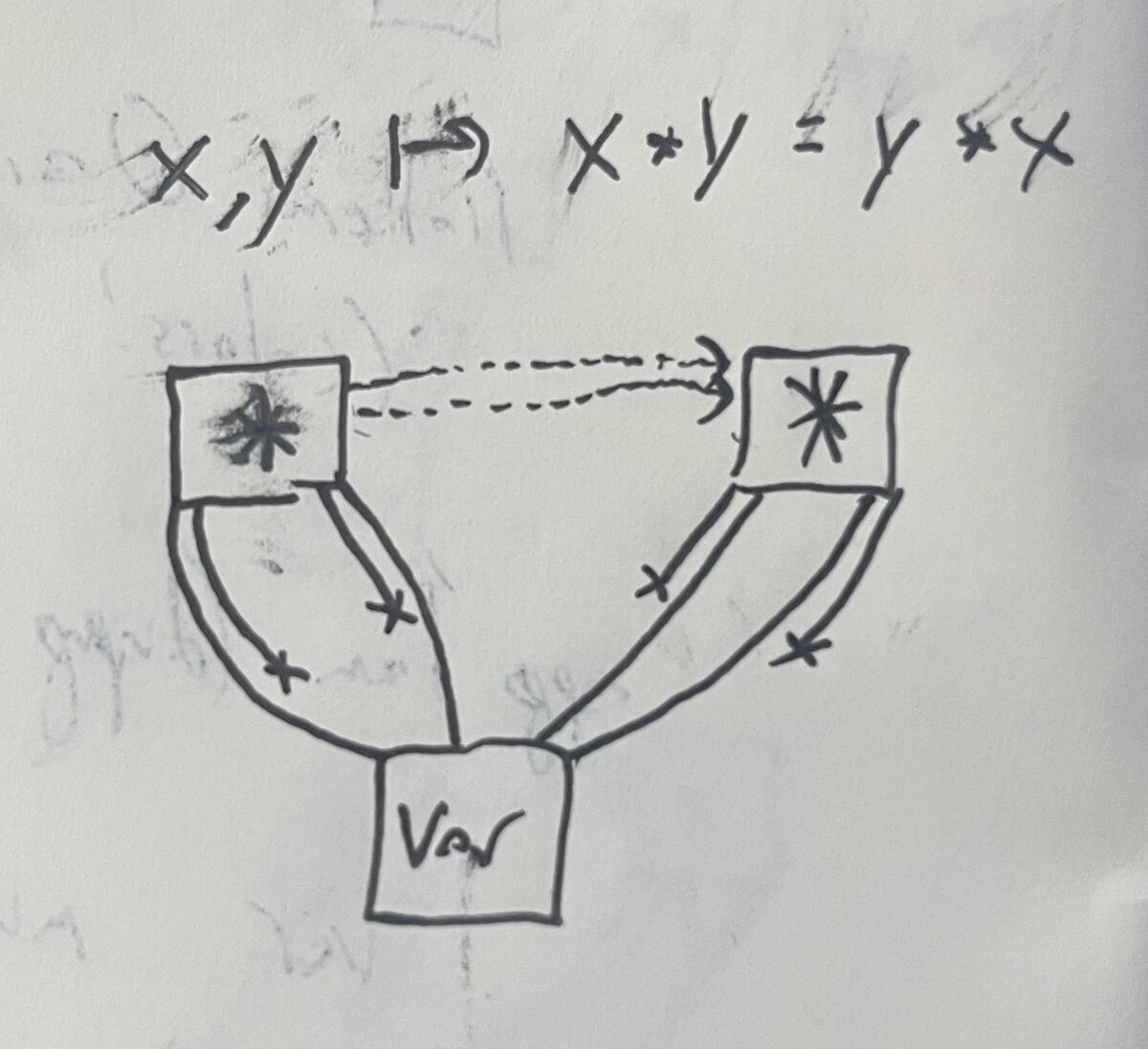}
\caption{ A visualization of a lifting e-graph via thickening all edges into thinnings. This lifting e-graph represents $\text{lift}_{10}(\text{var}) * \text{lift}_{01}(\text{var}) \rightarrow \text{lift}_{01}(\text{var}) * \text{lift}_{10}(\text{var})$
}
\label{fig:liftegraph}
\end{figure}

\section{E-Matching}

E-matching in the typical case proceeds by composing thinnings as you traverse into rules. This is justified as an application of the lift pushing rule $\text{lift}_t(f(e_1, e_2)) \rightarrow f(\text{lift}_t(e_1),\text{lift}_t(e_2) )$

$$
\begin{aligned}
\mathrm{lift}_i(e_1)
  &=_? f(?a, ?b)
  && \text{goal} \\
\mathrm{lift}_i(f(e_2,e_3))
  &=_? f(?a, ?b)
  && \text{nondet e-class expand } e_1 \rightarrow f(e_2,e_3) \\
f(\mathrm{lift}_i(e_2), \mathrm{lift}_i(e_3))
  &=_? f(?a, ?b)
  && \text{lift-pushing}\\
\Rightarrow \quad
?a &= \mathrm{lift}_i(e_2) \quad \land \quad
?b = \mathrm{lift}_i(e_3)
  && \text{decompose } f
\end{aligned}
$$

There is more complexity coming from redundancies in union find $e_{\text{child}} \rightarrow \text{lift}_k(e_{\text{parent}})$.

The simplest choice of dealing with these is to have the match fail when it encounters these. E-matching into these expressions will be rewriting in expressions that are using variables in a redundant way. This is not likely to be a useful place to be spending computation.

However, if one does want to match into these, it is possible. It corresponds to solving for thinnings in the equation $\text{lift}_i(e_\text{parent}) = \text{lift}_{?j}(\text{lift}_k(e_\text{parent}))$, which may have multiple solutions. This corresponds in the named representation to $(x,y \mapsto 0) =_? (x,y \mapsto ?a * 0)$ having both $?a = x$ and $?a = y$ as solutions.

\section{Related Work}

Slotted e-graphs \cite{slotted} are a mechanism for efficient handling of $\alpha$ equivalence in e-graphs. The e-graph described in this work can be seen as an approach to taking the slotted notion of redundancy as primary and removing the notion of permutative renaming. This is to some degree how the idea was developed. By making slots totally ordered and using a sparse representation of thinnings that lists the index of kept vars rather than the bitvector representation bring the two types of e-graph in closer alignment.

Both this work and slotted are related to the concept of Hashing modulo alpha equivalence \cite{hashalpha}, but it is not always clear how to take the ideas there and make them work in the highly shared situation of e-graphs.

Co de-Bruijn notation \cite{McBride_2018} is also a direct inspiration for the Thinning e-graph. Essentially the thinning e-graph is just an application of the Co de-Bruijn ideas in the e-graph setting.

The pointwise functional semantics is inspired by the semantics of judgements in type theory $\llbracket x : \mathbb{R}, y : \mathbb{R} \vdash \sin(x) : \mathbb{R} \rrbracket$ \cite{Hofmann_1997}. Indeed, the notation $\mapsto$ could be replaced throughout by $\vdash$, but this brings in new connotations. 

Weakening, Lifting, and Thinning are very similar concepts. There exists prior work on Explicit Weakening calculi \cite{Wadler_2024}\cite{DAVID_GUILLAUME_2001}.

There are de Bruijn methodologies that allow the $succ$ constructor of the variables to float inside of expressions rather than being tied to the variable \cite{kiselyov}\cite{pavel}. Every variable is associated with the $zero$ constructor then.

\section{Further Work}

The original intent of this work was to deal with bound variables. It has come as somewhat a surprise that there is an interesting system of variables to discuss before one talks about bindings. Bindings are an inessential feature. Binders like \lstinline|lam|, \lstinline|forall|, \lstinline|exists|, \lstinline|integrate|, and \lstinline|subst| are semantically higher-order "projection functions", in the sense that they remove a variable from the context $(\bR^n \rightarrow T) \rightarrow (\bR^{n-1} \rightarrow S)$. They have a different lift pulling rule / transfer function than pointwise application combinators, but the machinery required seems to be not that different.

The thinning union find is interesting in that it refutes a misunderstanding that group axioms are necessary for union find edge annotations. A better framework for generalizations is ongoing work.

The connection of type theory semantics to e-graphs is compelling as an avenue to both explicate them in terms of ordinary data structures, make e-graphs more usable for type theory applications, and as an exemplar of elegant handling of internalizable proof structures.

\section{Acknowledgements}
Many discussions with Rudi Schneider about slotted e-graphs have heavily informed this work. Commentary by Conor McBride on Mastodon about thinnings kicked off this particular branch of ideas. I'd also like to thank Max Willsey and Cody Roux for many interesting and relevant discussions.
\bibliographystyle{plain}
\bibliography{thinning}

\end{document}